\documentclass[aps, prl
, twocolumn
, nofootinbib 
, superscriptaddress
]{revtex4-2}

\usepackage{graphicx}
\usepackage{xcolor}
\usepackage{tikz}
\usepackage[caption=false]{subfig}
\usepackage{mathrsfs,mathtools}
\usepackage{physics,amssymb}
\usepackage{bm}
\usepackage{braket}
\usepackage{listings}
\usepackage{cases}
\usepackage{comment}
\usepackage{soul}
\usepackage{cancel}
\usepackage{cases}
\usepackage[utf8]{inputenc}
\usepackage{url}
\usepackage{longtable}
\usepackage[normalem]{ulem}
\usepackage{xspace}
\usepackage{acronym}
\usepackage[colorlinks=true
,urlcolor=DARKBLUE
,anchorcolor=DARKBLUE
,citecolor=DARKBLUE
,filecolor=DARKBLUE
,linkcolor=DARKBLUE
,menucolor=DARKBLUE
,linktocpage=true
,pdfproducer=medialab
,pdfa=true
]{hyperref}

\newcommand{\calD}{\mathcal{D}}

\newcommand{\calH}{\mathcal{H}}

\newcommand{\calO}{\mathcal{O}}

\newcommand{\bae}[1]{\begin{align} #1 \end{align}}

\definecolor{MONZA}{HTML}{CF000F}
\definecolor{DARKBLUE}{HTML}{00008b}
\definecolor{DARKMAGENTA}{HTML}{8b008b}

\def\apre{a_{\mathrm{pre}}}
\def\apos{a_{\mathrm{post}}}
\def\Hinf{\calH_{\ast}}
\def\aast{a_{\ast}}

\acrodef{EoM}{equation of motion}
\newacro{CMB}{cosmic microwave background}
\newacro{EFT}{effective field theory}
\newacro{SM}{Standard Model}
\newacro{UV}{ultraviolet}
\newacro{IR}{infrared}
\newacro{GR}{general relativity}
\newacro{FLRW}{Friedmann--Lema\^{i}tre--Robertson--Walker}
\newacro{EoS}{equation-of-state}
\newacro{CGPP}{cosmological gravitational particle production}
\newacro{GPP}{Gravitational particle production}
\newacro{GW}{gravitational wave}

\newcommand{\multi}[1]{\begin{multline} #1 \end{multline}}

\begin{document}
\title{\boldmath 
Universality classes of sharp gravitational production of light particles
}

\author{Kanta Ito}
\email{kanta.ito.clab@nagoya-u.jp}
\affiliation{Graduate School of Science, Nagoya University, Nagoya, 464-8602, Japan}
\author{Yusuke Mikura}
\email{ymikura@asiaa.sinica.edu.tw}
\affiliation{Academia Sinica Institute of Astronomy and Astrophysics (ASIAA),
\\
No. 1, Section 4, Roosevelt Road, Taipei 106319, Taiwan}
\author{Shuichiro Yokoyama} \email{shu@kmi.nagoya-u.ac.jp}
\affiliation{Kobayashi Maskawa Institute, Nagoya University, Aichi 464-8602, Japan}
\affiliation{
Department of Physics, Nagoya University, Nagoya, Aichi 464-8602, Japan
}
\affiliation{
  Kavli Institute for the Physics and Mathematics of the Universe (WPI),
  \\
  The University of Tokyo, Kashiwa, Chiba 277-8583, Japan}

\begin{abstract}
\noindent
We establish a classification of the gravitational production of light particles across sharp transitions connecting different asymptotic forms of the scale factor. A family of smooth scale factors with a transition duration $\Delta\eta$ is characterized by its regularity in the sharp-transition limit $\Delta\eta\to 0$, leading to three qualitatively different universality classes: $C^0$, $C^1$, and $C^2$-or-smoother. We find that the $C^0$ class exhibits a quadratic enhancement in the energy density of produced particles, $\rho\propto\Delta\eta^{-2}$, while the $C^1$ class shows a logarithmic enhancement, $\rho\propto\ln(1/\Delta\eta)$. No sharp-transition enhancement occurs for the $C^2$-or-smoother class. Using a specific $C^\infty$ model of the scale factor, we derive the energy density analytically and confirm both the predicted quadratic scaling and its coefficient by numerically solving the mode equation. These results identify the regularity of the scale factor in the sharp limit as the key property controlling the sharp-transition enhancement.
\end{abstract}

\maketitle

\paragraph{\bf Introduction.---}

Gravitational particle production
is a ubiquitous phenomenon in curved spacetime~\cite{Birrell:1982ix, Mukhanov:2007zz}. It provides an intriguing mechanism for producing particles since it operates even when produced particles have only gravitational interactions. In a cosmological context, \ac{CGPP} is caused by the breaking of the time translation symmetry and can be efficient in the early universe~\cite{Parker:1968mv,Sexl:1969ix,Parker:1969au,Grib:1969ruc,Zeldovich:1970si,Zeldovich:1971mw}. A variety of cosmological applications have been investigated, including the amplification of inflationary quantum fluctuations~\cite{Guth:1982ec,Hawking:1982cz,Starobinsky:1982ee,Bardeen:1983qw}, gravitational reheating~\cite{Ford:1986sy,Spokoiny:1993kt,Peebles:1998qn,Hashiba:2018iff}, dark matter~\cite{Turner:1987vd,Kuzmin:1998uv,Chung:1998zb,Chung:1998ua,Kuzmin:1998kk,Kolb:1998ki,Chung:2001cb,Ema:2018ucl,Hashiba:2018tbu,Li:2019ves,Ema:2019yrd,Hashiba:2019mzm,Kolb:2025wyj}, baryogenesis~\cite{Bassett:1997az,Hashiba:2019mzm,Enomoto:2021hfv}, and gravitational waves~\cite{deGarciaMaia:1993ck,Giovannini:1999bh,Tashiro:2003qp,Kunimitsu:2012xx,Ema:2015dka}. See Refs.~\cite{Ford:2021syk,Kolb:2023ydq} for recent reviews of \ac{CGPP}.

Among the various epochs during which \ac{CGPP} can occur, the transition from inflation to a subsequent phase is particularly important because the rapid change of the background can strongly violate adiabaticity and enhance particle production~\cite{Ford:1986sy}. A realistic transition occurs over a finite timescale, and the resulting particle abundance can depend sensitively on its duration and profile. \ac{CGPP} across finite-duration transitions has been investigated analytically and numerically for both heavy and light fields in Refs.~\cite{Ford:1986sy,Chun:2009yu,Glenz:2009zn,Nishi:2016wty,Artymowski:2017pua,Hashiba:2018iff,Li:2019ves}. 

When the transition timescale is shorter than the Hubble timescale, the rapid evolution of the scale factor can induce an enhancement in the energy density of gravitationally produced particles. For light particles, a logarithmic enhancement has been found both analytically and numerically in several transition realizations~\cite{Ford:1986sy,Chun:2009yu,Kunimitsu:2012xx,Nishi:2016wty}. In addition to this logarithmic enhancement, Ref.~\cite{Glenz:2009zn} found from numerical calculations that the energy density exhibits a quadratic scaling in a particular model. These results, however, have been obtained for specific transition profiles and asymptotic behaviors of the scale factor. A general criterion has not been established for identifying which properties of the transition determine these distinct scaling behaviors and, more generally, what types of sharp-transition enhancement can occur.

In this Letter, we classify the gravitational production of light particles according to the regularity of the scale factor in the sharp limit $\Hinf\Delta\eta\to0$, where $\Delta\eta$ denotes the transition duration with respect to the conformal time and $\Hinf$ is the conformal Hubble scale at the transition. The sensitivity of \ac{CGPP} to the regularity of the scale factor has previously been recognized~\cite{Chung:1998zb,Glenz:2009zn}. Here, we consider a family of transitions in which the scale factor is smooth for finite $\Delta\eta$ but approaches different $C^n$ regularities in the sharp limit. We find that there are three qualitatively different universality classes. Specifically, $C^0\setminus C^1$ scale factors yield the quadratic enhancement $\rho\propto\Delta\eta^{-2}$, while the logarithmic behavior arises for $C^1\setminus C^2$ scale factors. For $C^2$ or smoother scale factors, no corresponding enhancement occurs. These three classes are summarized as
\bae{
a^4\rho \propto
\begin{cases}
\Delta\eta^{-2} ~, & C^0 \setminus C^1 ~,
\\
\ln \left(\frac{1}{\Hinf \Delta\eta}\right) ~, & C^1 \setminus C^2 ~, 
\\
\text{no $\Delta\eta$ enhancement} ~, & C^2\ \mathrm{or\ smoother} ~.
\end{cases}
}
As an explicit example of the $C^0\setminus C^1$ class, we consider a specific $C^\infty$ transition from inflation to kination introduced in Ref.~\cite{Hashiba:2018iff}, derive the coefficient of the quadratic enhancement analytically, and confirm that the predicted energy density coincides with the numerical result.

\paragraph{\bf Universality classes.---}

We consider a scalar field nonminimally coupled to gravity, whose mass is much smaller than the Hubble scale at the transition. Neglecting the mass term, the canonically-normalized Fourier mode $\chi_k$ obeys
\bae{
\chi_k''+\left(k^2 - (6\xi+1) \frac{a''}{a}\right)\chi_k = 0 ~,
}
where $a$ is the scale factor and primes denote derivatives with respect to the conformal time $\eta$. The parameter $\xi$ is $-1/6$ for conformal coupling and $0$ for minimal coupling. The amount of produced particles is characterized by the Bogoliubov coefficient, given by~\cite{Parker:1969au,Zeldovich:1971mw,Ford:1986sy}
\bae{\label{eq:Bogolibov}
\beta_{k} \simeq -\frac{i}{2k} (6\xi+1)\int \frac{a''}{a} e^{-2ik\eta} \dd\eta ~,
}
with which the occupation number of mode $k$ is calculated as $n_{k} = |\beta_{k}|^2$. This shows that the time dependence of $a''/a$ acts as the source of \ac{CGPP}.

Without specifying a particular transition profile, a smooth transition can be described by
\bae{
a (\eta) = [1-S(x)] \apre (\eta) + S(x) \apos (\eta) ~,
}
where $x\coloneqq (\eta-\eta_\ast)/\Delta\eta$, and $\apre$ and $\apos$ denote the asymptotic scale factors before and after the transition, respectively. The outer scale factors, $\apre$ and $\apos$, are joined at $\eta=\eta_\ast$ in the sharp limit $\Hinf\Delta\eta\to0$. The transition function $S(x)$ is taken to be $C^\infty$ and to satisfy $ S(-\infty)=0$ and $S(+\infty)=1$. 

We assume that the outer scale factors, $\apre$ and $\apos$, evolve on the Hubble timescale, whereas the transition profile $S(x)$ varies on the much shorter timescale $\Delta\eta$. The outer scale factors then vary only slowly across the transition and can be expanded locally around $\eta=\eta_\ast$. Denoting the scale factor at $\eta_\ast$ by $\aast$, the difference between the two outer scale factors can be written as
\bae{\label{eq:difference-outer}
\frac{\apos (\eta) - \apre (\eta)}{\aast} = \sum_{n=0}^{\infty} \frac{\calD_n}{n!} (\eta-\eta_\ast)^n ~,
}
where $\calD_n$ denotes the discontinuity in the $n$th derivative of the scale factor at the sharp-transition point $\eta_\ast$,
\bae{
\calD_n \coloneqq \frac{\apos^{(n)} (\eta_\ast) - \apre^{(n)} (\eta_\ast)}{\aast} ~,
}
with $a^{(n)} (\eta)\coloneqq \dd^{n} a (\eta)/\dd \eta^n$.

We introduce the notion of $C^n$ regularity of the scale factor in the sharp limit. We say that a scale factor with a finite duration has $C^n$ regularity if its sharp limit satisfies
\bae{
\calD_{0} = \calD_{1} =\cdots =\calD_{n} = 0 ~, \quad
\calD_{n+1} \neq 0 ~.
}
For a scale factor with $C^n$ regularity, the leading difference between the outer scale factors~\eqref{eq:difference-outer} is proportional to $\calD_{n+1}$ as
\bae{
\frac{\apos (\eta) - \apre (\eta)}{\aast} \simeq \frac{\calD_{n+1}}{(n+1)!} (\eta-\eta_\ast)^{n+1} ~.
}
The quantity relevant to particle production, $a''/a$, can therefore be approximated as
\bae{\label{eq:adiabatic_contributions}
\frac{a'' (\eta)}{a (\eta)} \simeq \frac{\apre''(\eta)}{\aast} 
+ \calD_{n+1} \Delta\eta^{n-1} \Phi_n (x) ~,
}
where the transition profile is encoded in $\Phi_n$, defined by
\bae{
\Phi_n (x) \coloneqq \frac{\dd^2}{\dd x^2} \left[S(x) \frac{x^{n+1}}{(n+1)!}\right] ~.
}
 
We group the $C^n$ regularities into three classes: $n=0$, $n=1$, and $n\geq2$. Given the hierarchy between the Hubble and transition timescales, the first and second terms in eq.~\eqref{eq:adiabatic_contributions} scale differently as
\bae{\label{eq:order-estimate-1}
\frac{\apre'' (\eta)}{\aast} & \sim \calO(\Hinf^2) ~,
\\ \label{eq:order-estimate-2}
\calD_{n+1} \Delta\eta^{n-1} \Phi_n (x) & \sim \calO(\Hinf^{n+1}\Delta\eta^{n-1}) ~,
}
where $\calD_n \sim \calO(\Hinf^n)$ and $\Phi_n\sim \calO(1)$ within the transition. The relative magnitude of these two contributions therefore provides a natural criterion for this classification. We see the second term in eq.~\eqref{eq:adiabatic_contributions} dominates for $n=0$, the two contributions are comparable for $n=1$, and the contribution from the sharp transition is suppressed for $n \geq 2$.

\paragraph{\bf $\Delta\eta$ enhancement in the energy density.---}

Having defined the three classes, we examine the dependence of the energy density on the transition duration $\Delta\eta$. We assume that $a''/a$ in Eq.~\eqref{eq:Bogolibov} and its time derivatives vanish at asymptotic boundaries. After performing $n$ successive integrations by parts in Eq.~\eqref{eq:Bogolibov} for a scale factor with $C^n$ regularity and substituting the leading transition contribution in Eq.~\eqref{eq:adiabatic_contributions}, we obtain\footnote{For $n=1$, while the two terms in Eq.~\eqref{eq:adiabatic_contributions} are comparable in amplitude within the transition, they exhibit different behaviors in the Bogoliubov coefficient. The first term varies on the timescale $\Hinf^{-1}$, so that the corresponding Bogoliubov coefficient is suppressed for the relevant modes $\Hinf \ll k$. On the other hand, the second term varies on the shorter timescale $\Delta\eta$, which leads to a suppression only for $1/\Delta\eta \ll k$. We therefore retain only the sharp-transition contribution.}
\bae{\label{eq:Bogolibov-fourier}
\beta_k \simeq -\frac{i}{2k} (6\xi+1) \frac{\calD_{n+1}}{(2ik)^n} \hat{f}_n (2k\Delta\eta) ~,
}
up to an irrelevant phase. Here, we have introduced the function $f_n (x)$ and its Fourier transform $\hat{f}_n (q)$ by
\bae{
f_n (x) & \coloneqq \frac{\dd^n}{\dd x^n} \Phi_n (x) = \sum_{i=1}^{n+2} \binom{n+2}{i} \frac{x^{i-1}}{(i-1)!} S^{(i)} ~,
\\
\hat{f}_n (q) & \coloneqq \int f_{n} (x) e^{-i q x} \dd x ~.
}
Provided that the derivatives of the transition function satisfy
\bae{\label{eq:transition-conditions}
\lim_{x\to \pm \infty} \left[x^{i} \frac{\dd^{i}}{\dd x^{i}} S (x) \right] = 0 ~,
}
for $i = 1, \ldots, n+2$, the function $f_n (x)$ is localized around the transition and normalized as
\bae{\label{eq:f_integral}
 \int f_n (x)\dd x = 1 ~.
}
We note that Eq.~\eqref{eq:f_integral} implies $\hat{f}_n (0)=1$.

The $\Delta\eta$-dependent contribution to the energy density is governed by the subhorizon modes around the transition and is given by
\bae{
a^4 \rho \simeq \int_{\Hinf}^\infty \frac{k^3}{2\pi^2}|\beta_k|^2 \dd k ~.
}
Substituting the Bogoliubov coefficient~\eqref{eq:Bogolibov-fourier} and introducing $q\coloneqq2k\Delta\eta$, we obtain
\bae{\label{eq:energy-n}
    a^4\rho \simeq \frac{(6\xi+1)^2}{32\pi^2} \calD_{n+1}^2 \Delta\eta^{2(n-1)}\int_{2\Hinf\Delta\eta}^{\infty} q^{1-2n} |\hat{f}_n (q)|^2 \dd q ~,
}
where the dependence on the regularity class is explicit through $n$, while the detailed transition profile is encoded in $\hat f_n(q)$.

When $n=0$, the integrand of Eq.~\eqref{eq:energy-n} is regular at its lower endpoint, so its lower limit can be extended to zero at leading order in the sharp limit, giving
\bae{\label{eq:energy_0}
a^4 \rho \simeq \frac{(6\xi+1)^2}{32\pi^2} \calD_{1}^2 \Delta\eta^{-2}\int_{0}^{\infty} q |\hat{f}_0 (q)|^2 \dd q ~.
}
The remaining integral is a finite dimensionless quantity that depends on the transition profile but not on $\Delta\eta$. Thus, the $C^0$ class generically exhibits the quadratic sharp-transition enhancement $a^4\rho\propto\Delta\eta^{-2}$.
We note that the factor $q$ in the integrand suppresses the infrared part of the integral, while $e^{-iqx}$ in $\hat{f}_0 (q)$ suppresses the ultraviolet part.
The dominant contribution therefore arises from $q=\calO(1)$, or equivalently from $k=\calO(\Delta\eta^{-1})$.

Substituting $n=1$, the energy density becomes
\bae{
a^4\rho \simeq \frac{(6\xi+1)^2}{32\pi^2} \calD_2^2 \int_{2\Hinf\Delta\eta}^{\infty} q^{-1} |\hat f_1(q)|^2\dd q ~.
}
Because of the factor $q^{-1}$ in the integrand, the integral receives an enhanced contribution from the broad range $2\Hinf\Delta\eta\ll q\lesssim1$. Using $\hat f_1(0)=1$, the leading $\Delta\eta$-dependent contribution to the energy density is therefore
\bae{
a^4\rho \simeq \frac{(6\xi+1)^2}{32\pi^2} \calD_2^2 \ln\left(\frac{1}{\Hinf\Delta\eta}\right) ~.
}
Here, we have omitted a finite contribution independent of $\Delta\eta$, whose value generally depends on the details of the transition profile. Note that the overall amplitude of the logarithmic enhancement is set by $\calD_2$, while the numerical coefficient is independent of the detailed transition profile.

For the $C^2$ or smoother scale factors ($n \geq 2$), the integral~\eqref{eq:energy-n} is dominated by its lower limit and it scales as $\left(\Hinf\Delta\eta\right)^{2(1-n)}$. This cancels the explicit factor $\Delta\eta^{2(n-1)}$, so the resulting contribution becomes of order $\Hinf^4$ because of $\calD_{n+1} \sim \calO(\Hinf^{n+1})$. The numerical coefficient of the energy density depends on the full background evolution and cannot be determined from the sharp-transition contribution alone.

\paragraph{\bf Numerical verification.---}

We verify the predicted quadratic enhancement in the $C^0$ class using a specific $C^\infty$ transition profile introduced in Ref.~\cite{Hashiba:2018iff}, which models a smooth transition from inflation to kination. The scale factor is parametrized as
\multi{\label{eq:hashiba_yokoyama}
a^2(\eta) = \frac{1}{2}\left(1-\tanh{\frac{\eta}{\Delta\eta}}\right)\frac{1}{1+\Hinf^2\eta^2}
\\
+\frac{1}{2}\left(1+\tanh{\frac{\eta}{\Delta\eta}}\right)(1+\Hinf\eta) ~,
}
where the corresponding outer scale factors can be read as
\bae{
\apre = \frac{1}{\sqrt{1 + \Hinf^2 \eta^2}} ~, 
\quad 
\apos = \sqrt{1 + \Hinf \eta} ~.
}
In this model, the sharp-transition point is $\eta_\ast=0$, where the outer scale factors satisfy $\apre(0)=\apos(0) = \aast = 1$, whereas their first derivatives are $\apre'(0) = 0$ and $\apos'(0)= \Hinf/2$. Thus, the scale factor is continuous but its first derivative is discontinuous in the sharp limit $\Hinf\Delta\eta\to0$, indicating that the scale factor~\eqref{eq:hashiba_yokoyama} belongs to the $C^0$ class, with $\calD_1 = \Hinf/2$.

We derive the analytic expression of the energy density, including its coefficient. Since Eq.~\eqref{eq:hashiba_yokoyama} interpolates the squares of the two outer scale factors, we first relate its transition function to that introduced in our general parametrization. Defining
\bae{\label{eq:S_HJ_square}
S_{\rm HY}(x) = \frac12 \left[1 + \tanh \left(x\right)\right] ~,
}
Eq.~\eqref{eq:hashiba_yokoyama} can be written as 
\bae{
a^2 (\eta) = \left(1 - S_{\rm HY}(x) \right) \apre^2 (\eta) + S_{\rm HY} (x) \apos^2 (\eta) ~,
}
with $x=\eta/\Delta\eta$.
By contrast, our general parametrization interpolates the scale factors themselves as
\bae{
a (\eta) = \left(1 - S(x) \right) \apre (\eta) + S(x) \apos(\eta) ~.
} 
We notice that these two transition functions are algebraically related by
\bae{
S = S_{\rm HY} \frac{\apos + \apre}{a + \apre} \simeq S_{\rm HY} + \calO(\Hinf\Delta\eta) ~.
}
Thus, in the sharp limit, $S_{\rm HY}$ determines the leading transition profile for the scale factor itself. Using $S(x)=S_{\rm HY}(x)$ at leading order, the localized function $f_0 (x)$ takes the form
\bae{
f_0 (x) = \sech^2 (x) \left(1 - x \tanh{x}\right) ~,
}
which satisfies the boundary condition~\eqref{eq:transition-conditions}. The corresponding Fourier transform can be obtained in closed form as
\bae{
\hat{f}_0 (q) = \int \dd x f_0 (x) e^{-i q x} = \frac{\pi^2}{4} q^2 \frac{\cosh (\pi q /2)}{\sinh^2 (\pi q /2)} ~.
}
The profile-dependent integral can then be evaluated analytically as
\bae{
\int_{0}^{\infty} q |\hat{f}_0 (q)|^2 \dd q = \frac{10 (2\zeta(3)+\zeta(5))}{\pi^2} ~,
}
where $\zeta(s)$ is the Riemann zeta function. Substituting this and $\calD_1 = \Hinf/2$ into Eq.~\eqref{eq:energy_0}, we obtain the analytic prediction of the energy density as
\bae{
a^4\rho \simeq \frac{5 (2\zeta(3)+\zeta(5))}{64\pi^4} (6\xi+1)^2 \frac{\Hinf^4}{(\Hinf\Delta\eta)^2} ~.
\label{eq:analytic}
}

\begin{figure}
    \centering
\includegraphics[width=\columnwidth]{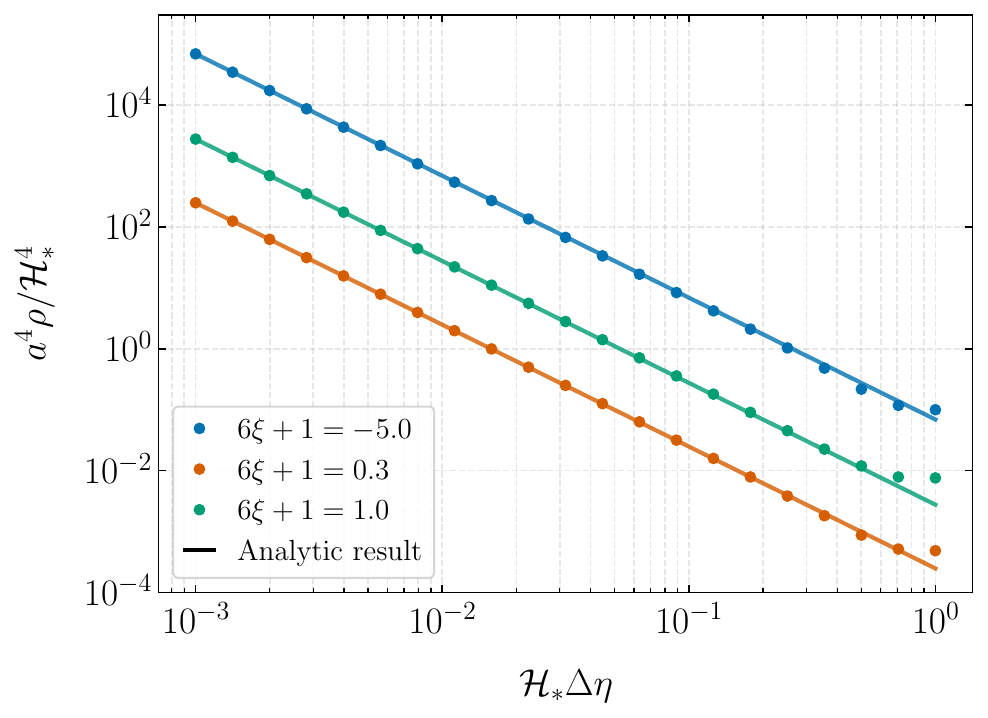}
    \caption{
    Numerical test of the $C^0$ sharp-limit prediction for the energy density using the $C^\infty$ scale factor given in Eq.~\eqref{eq:hashiba_yokoyama}. The points are obtained by solving the mode equation numerically, while the solid lines show the analytic predictions~\eqref{eq:analytic}. The numerical results agree with the analytic predictions for $\Hinf\Delta\eta\ll1$.
    }
\label{fig:energy_density}
\end{figure}

We compare the analytic result with a numerical calculation obtained by solving the mode equation. To this end, we need to specify the adiabatic vacua in the asymptotic inflationary and kination phases. During inflation, the corresponding adiabatic vacuum is known as the Bunch--Davies vacuum~\cite{Bunch:1978yq}, given by
\bae{\label{eq:BD_vacuum}
\chi_k^\mathrm{inf} (\eta) = \frac{\sqrt{-\pi\eta}}{2}e^{i\pi(2\nu_\mathrm{inf}+1)/4} H^{(1)}_{\nu_{\mathrm{inf}}}(-k\eta) ~,
}
where $H^{(1)}_{\nu_{\mathrm{inf}}}$ is the Hankel function of the first kind with index $\nu_\mathrm{inf} \coloneqq \sqrt{9+48\xi}/2$.
During kination, the adiabatic vacuum instead takes the form
\bae{\label{eq:kination_vacuum}
\chi_k^{\mathrm{kin}}(\eta) = \frac{\sqrt{\pi\eta}}{2} e^{-i\pi(2\nu_\mathrm{kin}+1)/4}H^{(2)}_{\nu_\mathrm{kin}}(k\eta) ~,
}
with $\nu_\mathrm{kin}\coloneqq\sqrt{-3\xi/2}$.
Using the Bunch--Davies vacuum~\eqref{eq:BD_vacuum} as the initial condition, we evolve the mode function $\chi_k$ across the transition and extract the Bogoliubov coefficient in the asymptotic kination phase as
\bae{\label{eq:bogoliubov-numerical}
\beta_k = i \left(\chi_k \chi^{\mathrm{kin}\prime}_k - \chi^{\prime}_k \chi^{\mathrm{kin}}_k \right) ~.
}
Fig.~\ref{fig:energy_density} shows the resulting energy density as a function of the transition duration.\footnote{To focus on \ac{CGPP}, we consider a parameter range $- 15 < 6\xi + 1 < 15/8$ in which no tachyonic instability occurs~\cite{Chakraborty:2024rgl}.}
As the transition becomes sharper, the numerical result approaches the analytic prediction. The agreement in the sharp-transition regime confirms both the predicted quadratic scaling and its analytic coefficient. Once $\Hinf\Delta\eta$ is close to unity, the deviation appears because the sharp-limit expansion is no longer applicable.

\paragraph{\bf Conclusions.---}

We have established the sharp-limit classification of the gravitational production of light particles across finite-duration transitions. Although the scale factor is $C^\infty$ for nonzero $\Delta\eta$, the $\Delta\eta$ dependence of the energy density is controlled by the regularity of the scale factor in the sharp limit $\Hinf\Delta\eta\to0$. There are three qualitatively different universality classes. The $C^0$ class exhibits the quadratic enhancement
\bae{
a^4\rho\propto\Delta\eta^{-2} ~,
}
while the $C^1$ class shows the logarithmic enhancement
\bae{
a^4\rho\propto \ln\left(\frac{1}{\mathcal \Hinf\Delta\eta}\right) ~.
}
No sharp-transition enhancement occurs for the $C^2$-or-smoother class. The previously studied quadratic enhancement~\cite{Glenz:2009zn} and logarithmic enhancement~\cite{Ford:1986sy,Chun:2009yu,Kunimitsu:2012xx,Nishi:2016wty} are thus identified as concrete realizations of the $C^0$ and $C^1$ classes, respectively. Using the benchmark smooth transition, we have further verified the $C^0$ prediction, including its analytic coefficient, by numerically solving the mode equation. These results demonstrate that the regularity of the scale factor in the sharp limit provides an organizing principle for \ac{CGPP} across sharp but smooth transitions.

The strong particle production in the $C^0$ class can have important phenomenological consequences. An immediate application is gravitational reheating after inflation. The quadratic enhancement, stronger than the logarithmic, can substantially increase the radiation abundance immediately after inflation and thereby shorten a subsequent kination phase. This is relevant to the constraints from big bang nucleosynthesis on the energy density of the primordial gravitational-wave background~\cite{Gouttenoire:2021jhk}. 
Our classification may also leave characteristic imprints on the spectrum of the primordial gravitational waves. Since each polarization of the primordial gravitational waves obeys the same mode equation as a massless minimally coupled scalar field, the $C^n$ regularity also affects its spectral shape. A detailed analysis of its scaling and amplitude is left for future work.

\acknowledgments

S.Y. is supported by JSPS KAKENHI Grant Numbers JP23H00108, and JP24K00627.

\bibliography{bib_submit}

\end{document}